\documentclass[twocolumn,showpacs,preprintnumbers,amsmath,amssymb]{revtex4}
\usepackage{graphicx}
\usepackage{dcolumn}
\usepackage{bm}
   
\begin{document}
\title{Elastic consequences of a single plastic event :\\ 
a step towards the microscopic modeling of the flow of yield stress fluids.}

\author{Guillemette Picard$^{1}$, Armand Ajdari$^{1*}$, Fran\c{c}ois Lequeux$^{2}$, Lyd\'eric Bocquet$^{3}$}
\address{$^{1}$Laboratoire de Physico-Chimie Th\'eorique, UMR CNRS 7083,\\
 $^{2}$ Laboratoire de Physico-Chimie Macromol\'eculaire, UMR CNRS 7615,\\
{\rm both at} ESPCI, 10 rue Vauquelin, F-75005 Paris, France\\
$^{3}$ Laboratoire PMCN, UMR CNRS 5586, Universit\'e Lyon I, 43 Bd du 11 Novembre 1918, 69622 Villeurbanne Cedex, France
}
\date{Submitted to EPJE, January 16 2004}
  
\begin{abstract}
With the eventual aim of describing flowing elasto-plastic materials, we focus on the elementary brick of such a flow, a plastic event, and compute the long-range perturbation it elastically induces 
in a medium submitted to a global shear strain. We characterize the effect of a nearby wall on this perturbation, and quantify the importance of finite size effects. Although for the
sake of simplicity most of our explicit formulae deal with a 2D situation,
our statements hold for 3D situations as well. 
\end{abstract}
\maketitle
\section{Introduction}

 An increasing body of experiments on macroscopic flow of various complex systems evidence spatially heterogeneous behaviour.  We focus here on the (large) sub-class of such systems that display a macroscopic yield stress (among which
foams, suspensions, emulsions, colloidal glasses,...). Typically these
systems flow homogeneously at large stress/shear rate,
whereas at intermediate shear rate they may exhibit spatial coexistence between a flowing and a frozen region \cite{zuk,pignon,varnik,cous,deb} or intermittent heterogeneous flow \cite{pignon}.  
To this point, there is little insight as to whether the mechanisms leading to such macroscopic behaviours are generic or dependent on the specific microscopic structure of the fluid and the corresponding interactions.

A general class of  ``elasto-plastic''  models has been 
put forward to apprehend those macroscopic 
behaviour, which was
first applied to seismologic modelisation \cite{chen}.  
In these models, the medium first responds elastically to a global forcing (either stress or strain). The deformation or stress can then locally induce a rearrangement or {\it plastic event}, if a local threshold is reached. Such a plastic event locally relaxes a stress that is elastically redistributed in the medium, and can trigger other local events.  In this picture, the macroscopic flow is 
the outcome of the collectively organized sequence of local rearrangements.  
Although this mesoscopic description seems very reasonable,  
many questions remain to be answered
for this scenario to be operational.
First, what is (are) the basic plastic event(s), and how can it (they) be 
identified in a given flowing complex material ?
Second, what is the constitutive (dynamic) equation that 
describes such a single plastic event under a local forcing ? 
Third, how does such an "event", locally relaxing stress, perturb the 
surrounding medium ? The answer to this last question is 
obviously linked to the nature of the plastic event. 

As to the first two questions, {\it i.e.} the nature and the description of the plastic event, various convincing pictures
have been proposed in the literature. In a pioneering work, 
Bulatov and Argon introduced a phenomenological description of a single plastic event, which allowed them to describe
many properties of macroscopic plastic flows \cite{bulatov1,bulatov2,bulatov3}. In their simulation, the unit cell can undergo several fixed plastic deformations specific to their hexagonal geometry.
Later, on the basis of molecular simulations of a Lennard Jones glass under imposed shear stress \cite{langer2}, and building on earlier works by Spaepen and Argon,
Falk and Langer introduced the notion of shear transformation zone (STZ), which described a local limited zone where rearrangements occur. 
The occurrence of very localized plastic events is 
most easily evidenced in foams \cite{deb}, 
where they take the form of T1 rearrangements.
Langer \cite{langer1} then constructed an analytical "mean field" 
elasto-plastic model, introducing STZ as zones with a plastic tensorial deformation.
More recently, Baret {\it et. al.} \cite{baret}, and Braun \cite{braun} performed numerical simulations on lattices, in
which a plastic event consists in a local scalar displacement occurring when the local stress reaches a yield stress.
Thus, within a very general class of elasto-plastic model, 
the notion and the description of a plastic event
is now well documented and clarified. 
However, a clear description of the consequences of a localized
plastic event on the stress distribution in the material (third question)
still needs to be constructed. 

This is the purpose of the present paper : 
using a rather general description for a localized plastic event, 
we compute the long-range {\it elastic} perturbation that such an event 
induces in an elastic material. We characterize its symmetry and amplitude,
as well as the way it is modified if the event occurs close
to a solid boundary. An {\it a priori} counter-intuitive result 
which emerges from our calculations, is
the crucial role played by finite size effects in the modeling of flowing elasto-plastic materials.
We limit ourselves here to the study of the elastic effects of a single event, 
and leave for a later report the analysis of the 
collective organization of the plastic events when the material is 
flowing.

The paper is organized as follows. In section II we specify 
the general ``elasto-plastic" model that we use: we assume that the medium is homogeneous and isotropic, as well as incompressible for simplicity. 
In section \ref{sec_infinite} we consider an infinite geometry,
and describe a local plastic event induced by shearing 
and the full characterization of the perturbation it elastically induces.
In this case there is no difference between a forcing at imposed stress or 
imposed strain.
In section \ref{sec_finite}, we focus on finite size geometries, 
where the system is bounded by solid walls. First, we describe how a wall attenuates the perturbation induced by an event occurring in its vicinity. Secondly, we explicit 
how in a finite size medium, the perturbation depends on the global forcing. We 
calculate the average stress relaxation induced by an event at imposed strain,
and give explicit formulae to compute the corresponding stress field relaxation everywhere. In section \ref{sec_conseq}, we conclude and briefly highlight important consequences for the modeling of flowing systems.

\section{Elasto-plastic model}
\label{sec_elastop}

We assume, following many of the previously quoted studies,
that the displacements and deformations are given by the simple superimposition
of a plastic flow (the localized plastic events) and an elastic 
distortion of the medium. We further assume
that the medium is homogeneous, isotropic, and linearly elastic. 
In addition, we focus for simplicity on the incompressible case (the compressible case can be studied following the same lines), so that the elastic properties
of the medium are fully described by the shear modulus $\mu$.

Denoting ${\bf u}({\bf r})$
the total displacement vector at position $\bf{ r}$, the strain tensor 
is given by ${\bm \epsilon}=\frac{1}{2} (\nabla {\bf u} + (\nabla {\bf u})^{t})$.
From our hypotheses, this 
total strain is the sum of an elastic
strain and a plastic strain (non-zero only at the locus of plastic events):
\begin{equation}
\label{toteps}
{\bm \epsilon} = {\bm \epsilon}^{pl} + {\bm \epsilon}^{el}
\end{equation}
Incompressibility corresponds to :
\begin{equation}
\label{incomp} 
\nabla. {\bm u} = 0
\end{equation}

With the hypotheses of linear elasticity and incompressibility, 
the total stress tensor is ${\bf s} = -p {\bf 1} + {\bm \sigma}$ where $p$ is the pressure and $ \sigma $ verifies : 
\begin{equation}
\label{elasticity}
{\bm \sigma} = 2 \mu {\bm \epsilon}^{el}=2 \mu
{\bm \epsilon}-2 \mu {\bm \epsilon}^{pl}
\end{equation}
As we have in mind the slow flow of pasty materials we neglect inertial effects
so that mechanical equilibrium simply requires :
\begin{equation}
\label{mech_eq}
 \nabla . ({\bm \sigma} -p {\bf 1}) =0
\end{equation}
We consider the classical situation where an applied shear
(either imposed deformation or stress)
induces elastic loading of the material, up to the point
where it triggers a single localized plastic event.
The consequent state of the medium in response to the applied forcing is 
here the sum of a purely elastic response to the forcing (denoted with superscripts $0$) and of the perturbation induced by the occurrence
of the plastic event (denoted with superscripts $1$).

With these notations :
\begin{eqnarray}
{\bm \epsilon} &=& {\bm \epsilon}^0 + {\bm \epsilon}^1 \nonumber \\
{\bm \epsilon}^{el} = {\bm \epsilon}^{el 0}+ {\bm \epsilon}^{el 1} &;& {\bm \epsilon}^{pl} = {\bm \epsilon}^{pl 1} \nonumber \\
{\bm \sigma} &=&{\bm \sigma}^{0} +{\bm \sigma}^{1}
\end{eqnarray}
where ${\bm \epsilon}^{pl 1}$ describes the localized plastic event.
To go any further we need to pay attention to the boundary
conditions imposed on those fields, which brings us 
to specify the global geometry of the system.
We start below with an infinite medium, 
before addressing finite-size effects in section IV.

\section{Infinite medium}
\label{sec_infinite}

We start with the limit case of an infinite medium,
where a driving at infinity imposes either an applied
shear strain or an applied shear stress.
The purely elastic response ${\bm \epsilon}^{el0}$
is homogeneous. Let us then focus on the perturbation $1$
generated by a plastic event described by ${\bm \epsilon}^{pl}$.  
If the system is stress driven, the boundary condition 
for the induced perturbation is:
\begin{equation}
\label{bcinfty_stress}
{\bf \sigma}^1(\infty) \rightarrow 0
\end{equation}
whereas for a strain controlled system, the boundary condition reads :
\begin{equation}
\label{bcinfty_gam}
{\bf u}^1(\infty) \rightarrow 0
\end{equation}

Equations (\ref{incomp}), (\ref{elasticity}) and (\ref{mech_eq}) can then be reformulated in a well defined problem for the perturbation field:
\begin{eqnarray}
\label{source_pl}
\nabla. {\bf u}^1 =0 \nonumber \\
 \nabla. (2 \mu {\bf \epsilon} ^1  -p^1 {\bf I}) =2 \mu \nabla . 
{\bm \epsilon}^{pl}
\end{eqnarray}
which must be solved either with either (\ref{bcinfty_stress}) or (\ref{bcinfty_gam}). 

Obviously the problem at hand is directly related to 
the response of a purely elastic (incompressible, isotropic, homogeneous) system to a punctual force.
Let a force  ${\bf f} $ act on the medium at the position $\bf{r'}$. The displacement field $ {\bf u}$ at position ${\bf r}$ created by this force is given by:
\begin{eqnarray}
\label{eq_oseen}
\nabla. {\bf u} =0 \nonumber \\
 \nabla. (2 \mu {\bf \epsilon} -p {\bf I}) + {\bf f} \delta({\bf r})= 0
\end{eqnarray}
The solution of this system, that actually satisfies
simultaneously both types of boundary condition (${\bf \sigma} (\infty) \rightarrow 0$ and ${\bf u}(\infty) \rightarrow 0$), is the Oseen tensor which is most
easily dealt with in reciprocal space (we use hats to denote Fourier transforms):
\begin{eqnarray}
\hat{{\bf u}}({\bf q})& =& \hat{{\bf O}} ({\bf q})  . {\bf f} \nonumber
\end{eqnarray}
with:
\begin{eqnarray}
\hat{\bf O}({\bf q}) &=&  \frac{1}{ \mu q^2} ({\bf 1} -\frac{{\bf q}{\bf q}}{q^2}) \nonumber \\
\end{eqnarray}  

With this tool we can construct the solution to the system (\ref{source_pl}).
This solution is clearly 
also independent of the specific boundary condition (6) or (7),
and can be simply written:
\begin{equation}
\label{div_epspl}
{\hat {\bf u}}^{1}({\bf q}) =2 \mu \hat{{\bf O}} ({\bf q})  .(i{\bf q}.\hat{{\bm \epsilon}^{pl}})
\end{equation}
At this point, no assumption has been made on the nature of the plastic event
(fully described here by the localized ${\bm \epsilon}^{pl}({\bf r})$,
or equivalently by $\hat{{\bm \epsilon}^{pl}}({\bf q})$). 
Therefore this expression gives 
in a very general form the displacement field induced 
by a plastic strain in an infinite medium (for both type of boundary conditions). 
By derivation and using (3) one easily obtains a similar formula
in reciprocal space for the stress perturbation $\sigma^1$. Both relations
lead in real space to expressions for the propagators for displacement and stress 
respectively. Rather than producing these formulae in a formal general
context we specify to a given geometry first.\\

\subsection{Plastic events with a simple shear symmetry in 2D}

\begin{figure}[t]
\includegraphics[width=7cm]{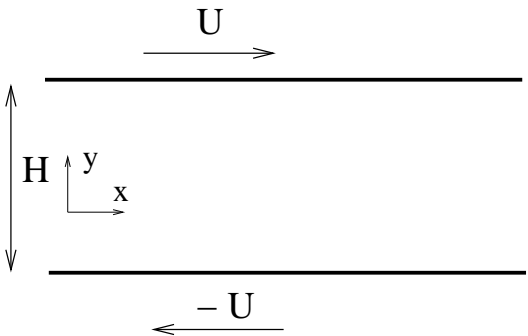}
\caption{The medium is sheared in the $y$ direction at constant strain $\gamma$. 
the infinite system of section III corresponds to $ H \rightarrow \infty, U \rightarrow  \infty $ and $\frac {U}{H} = \gamma/2 $}
\label{fig : cis2d_inf}
\end{figure}

To pursue analytically without dealing with opaque tensorial formulae, 
{\em we now make the simplifying assumption that the local plastic event has the symmetry of the global forcing which we chose to be that of simple shear}.
We further focus on the two dimensional case : 
the $H \rightarrow \infty$ limit of Fig. \ref{fig : cis2d_inf}. 
We stress however that most our conclusions
are also valid for the 3D situation (see comments further).

With the hypothesis above, the plastic deformation tensor corresponds to simple shear $\epsilon_{xy}^{pl}=\epsilon_{yx}^{pl}$ , and $\epsilon_{xx}^{pl},\epsilon_{yy}^{pl}$ are neglected.
Expression (\ref{div_epspl}) becomes:
\begin{eqnarray}
\label{dpt_gam}
{\hat  u}_{x}^{1}({\bf q}) &=& 2 \mu( \hat{O}_{xx} . i q_y + \hat{O}_{xy}. i q_ x)\hat{\epsilon}^{pl}_{xy} \nonumber  \\
{\hat  u}_y^{1}({\bf q}) &=& 2 \mu( \hat{O}_{xy}. i q_y + \hat{O}_{yy}. i q_x) \hat{\epsilon}^{pl}_{xy}\nonumber \\ 
\end{eqnarray}

\begin{figure}[b]

\includegraphics[width=6cm]{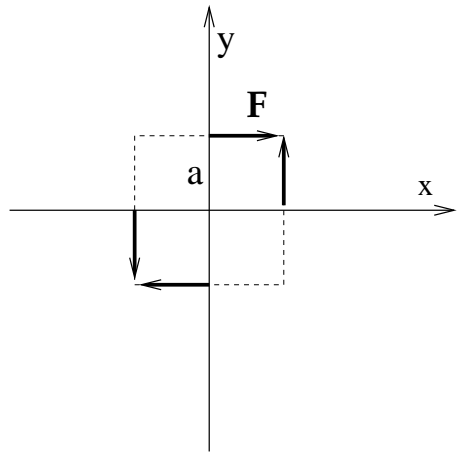}
\caption{The perturbation due to a localized plastic shear is equivalent to the perturbation due to a set of two dipoles of forces  with $F =a \mu \epsilon_0$.}
\label{fig : quadrup}
\end{figure}

For a localized plastic event $\epsilon_{xy}^{pl}=\epsilon_0a^2\delta({\bf r})$
(i.e. of typical amplitude $\epsilon_0$ and microscopic spatial extent $a^2$),
the perturbation displacement described by (\ref{dpt_gam}) is analogous to the displacement induced by a set of two dipoles of forces $({\bf F},2a) $ as 
represented on figure \ref{fig : quadrup}, with $F = a \mu \epsilon _0$
(or more precisely its limit for $a \rightarrow 0$ with $aF=\mu\epsilon_0 a^2$ kept
constant).

The shear stress perturbation corresponding to (\ref {dpt_gam}) can be obtained using (3), and reads : 
\begin{eqnarray}
\hat {\sigma}_{xy}^{1} &=&2 \mu^2 (q_y^2 \hat{O}_{xx}+ q_x^2 \hat{O}_{yy} +2  q_x q_y \hat{O}_{xy}) \hat {\epsilon}_{xy}^{pl} \nonumber \\
&-& 2 \mu \hat {\epsilon}_{xy}^{pl}
\label{oseen4p}
\end{eqnarray}
Somewhat similar formulae can be obtained for $\sigma_{xx}$ and $\sigma_{yy}$
but we will in the following mostly focus on the shear stress.

We define formally the {\it propagators} ${\bf P}^{\infty}$, $G^{\infty}$ that describe 
the consequences in terms of displacement and elastic shear stress 
of a single plastic event in an infinite medium by:
\begin{eqnarray}
\label{propag_inf}
{\bf u}^1 ({\bf r})&=&\int\!\! d{\bf r'} {\bf P}^{\infty}({\bf r-r'}) 
\epsilon_{xy}^{pl} ({\bf r'}) \ \\
 \sigma_{xy}^1 ({\bf r})  &=& 2 \mu \int\!\! d{\bf r'} 
G^{\infty}({\bf r-r'})
\epsilon_{xy}^{pl} ({\bf r'})
\end{eqnarray}
Practically, the propagator for the displacement field can be derived from equation Eq. (\ref{dpt_gam}) either with the Fourier Transform or by derivation in real space of the Oseen tensor.  The propagator for the shear stress is then deduced from linear elasticity.

\begin{figure}[t]
\includegraphics[width=7cm]{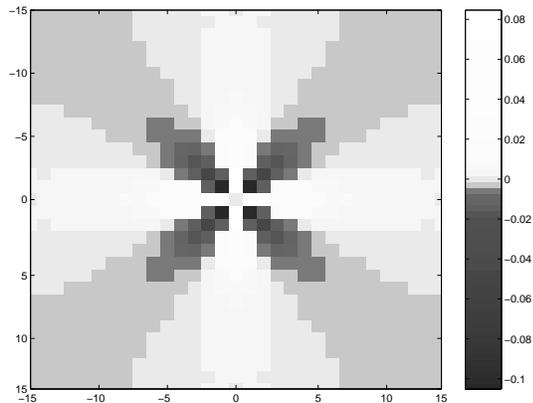}
\caption{Perturbation of the shear stress field for a plastic event 
occurring at the origin in a medium  submitted to a shear strain or stress.}
\label{lobe}
\end{figure}

In the present two-dimensional geometry, explicit formulae in reciprocal and real space are:
\begin{eqnarray}
\hat{G}^{\infty}&=&  -4 \frac{q_x^2 q_y^2}{q^4} \\
G^{\infty} (r,\theta) &=& \frac{1}{\pi} \frac{2}{r^2} cos(4 \theta) 
\label{g4p2d}
\end{eqnarray}
Hence,  in a system forced with a symmetry of simple shear, 
the perturbation of the shear stress due to a localized plastic event is of quadrupolar symmetry (see Fig. \ref{lobe}) and decreases with a power law $\frac{1}{r^2}$  in two dimensions (in three dimensional systems, the quadrupolar symmetry is conserved and the power law  is $\frac{1}{r^3}$).

\subsection{Global effect of a plastic event :}

We now consider more global effects of the same localized plastic event
$\hat{\epsilon}_{xy}^{pl}({\bm r'}) = a^2 \epsilon_0 \delta({\bf r'})$, 
namely quantities integrated over a whole layer (of constant $y$).
The following integral rules are easily derived from the previous 
calculations:
\begin{eqnarray}
\int^{+ \infty}_{-\infty} \sigma_{xy}^{1} (x,y) dx &=& 0 \nonumber \\
\int^{+ \infty}_{-\infty} u^1_x (x,y) dx &=&  \mbox{Sign}(y) a^2 \epsilon_0 \nonumber \\
\int^{+ \infty}_{-\infty} u^1_y (x,y) dx &=&0
\label{eq_sum}
\end{eqnarray}
The first equation states that the shear stress 
resulting from the plastic event is redistributed in such a way 
that the integrated stress on every layer is unchanged,
i.e. there is no net release of stress over a layer, and consequently no change in the net force applied from above on the system!
The second relation indicates that the average horizontal displacement in a
(horizontal) layer depends only on whether it is above or below the event 
(but not on its distance to the event), while the third equation
expresses that the average vertical displacement over a layer is zero.
For simple shear in three dimensions, similar equations hold
for quantities integrated over planes perpendicular to the loading
displacement gradient.

\subsection{Relation to other studies}
\label{sec_discevent}

Let us now compare the results we have obtained at the end of section A,
for the displacement and stress fields induced by a single localized plastic
event  of simple shear symmetry, to related studies in the literature.


First, our results can be compared with the full analytical description of an elasto-plastic inclusion in an elastic matrix by Eshelby \cite{eshelby}. In that study, plastic shear strain occurs only within the finite-sized inclusion
which yields a perturbation of the shear stress around the inclusion.
Whatever the explicit shape of the inclusion,
the long-range behaviour of that perturbation is strictly identical to the equivalent for the 3D case of expression (\ref{propag_inf})(as we have checked).

To study collective effects, Baret {\it et al.} \cite{baret} simulated 
elements with local yield stresses on a 2D-lattice (semi-periodic boundary conditions). In contrast with the present study, 
they modeled the plastic event by a simple scalar displacement. 
As a check, we computed the plastic strain 
tensor corresponding to such a local event in expression (\ref{div_epspl}), and calculated the corresponding perturbed shear stress. 
This yields a perturbation of dipolar symmetry, consistent with the propagator  that they numerically evaluated on their lattice. This provides a
validation of our procedure, but mostly underlines that the nature and symmetry 
of the elementary plastic event seriously affect the propagator describing its consequences, and therefore potentially the collective interplay of such events
and the resulting macroscopic flow behaviour. We believe that the form of
plastic event used in the present study is more suited for the actual description
of the flow of elastoplastic materials.

Kabla and Debr\'egeas \cite{kabla} performed an explicit 
numerical simulation of a two dimensional foam under shear strain. 
In their quasi-static procedure, at each step the length of the film 
is minimized at constant bubble volume. 
They mimic 'T1' event by reorganizing sets of four bubbles 
when film length decreases below a critical value. 
They studied the stress rearrangements following such T1 events in their
simulation. Averaging over many such events, they found that statistically
these stress perturbations have a quadrupolar symmetry (
with a slight tilt of the axes with respect to those of the macroscopic shear
({\it x,y}), probably due to a structuration of the foam by the flow into a slightly non-isotropic medium). They observed that this stress field coincides with that generated in an elastic medium by a set of dipoles with the orientation of the global forcing. What is remarkable is that the outcome of their cellular simulation
is very consistent with the outcome of the (elastic) continuum approach followed
here. 

To describe the deformation of plastic amorphous materials,
Langer has introduced the concept of  introduces shear transformation zones (STZ).
In \cite{langer1}, he studies the response of a 2D material to an applied
deviatoric stress. The plastic strain tensor is described without 
any assumption on its orientation. The results in that paper 
suggest that for a plastic strain tensor with the symmetry of the forcing, the deviatoric stress induced by the STZ is of quadrupolar symmetry, with a power law 
decrease $\frac{1}{r^4}$. Within our model, we find for such
an event also a quadrupolar symmetry but a power law decrease of $\frac{1}{r^2}$,  and do not understand this discrepancy.

We now turn to the most important part of our paper,
namely the effect of the finite-size geometry on the stress
distortion due to a localized event.

\section{Finite medium}
\label{sec_finite}

As in the analysis above, we consider 
here localized plastic events with the symmetry of the forcing (simple shear),
equivalent to a set of two dipoles of forces.
Therefore the propagator for such a plastic event can in principle
be viewed as the sum of the propagators for the four forces,
which brings us back to considerations pertaining to the Green's function 
for the effect of a single force on the elastic {\em finite} medium.

\begin{figure}[t]
\includegraphics[width=8cm]{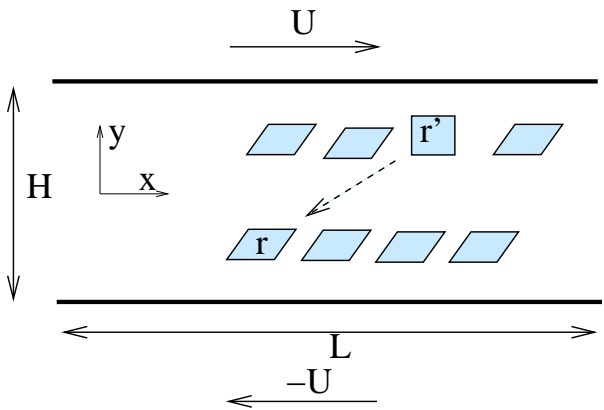}
\caption{ A shear strain $\frac{2U}{H}$ is applied to a two dimensional elasto-plastic medium. A plastic event occurs at position ${\bf r'}$ and is here represented by a square, and we seek for the elastic deformation
of the rest of the medium that will add up on the elastic loading
symbolized by the parallelograms.}
\label{cis2d_fini}
\end{figure}

We focus on the case of an imposed shear strain 
represented on figure \ref{cis2d_fini}, where two solid walls adhering
perfectly to the medium are shifted horizontally.

Therefore the boundary condition for the total displacement field is:
\begin{equation}
{\bf u}(x,\pm H/2)=\pm {\bf U } 
\end{equation}
We proceed with the same decomposition as in the infinite medium case. The displacement field is the sum of the homogeneous elastic loading and 
of the perturbation due to the plastic event (again we denote by ${\bf u}^1, 
{\bf \sigma}^1$ the displacement and deviatoric stress tensor 
induced by the plastic event). The boundary condition for the perturbation in 
the imposed strain regime is:
\begin{equation}
\label{bc_finite}
{\bf u}^1(x, \pm H/2)={\bf 0}
\end{equation}
The total response of the elasto-plastic medium is then :
\begin{eqnarray}
u_x(x,y) &=&  \frac{2 U y}{H} + u^1_x \nonumber \\
u_y(x,y) &=&  u^1_y \nonumber \\
\sigma_{xy}(x,y)&=& \frac{2 \mu U}{H} + \sigma^1_{xy}(x,y) \nonumber \\
\end{eqnarray}

\subsection{Wall effects}
\label{sec_wall}

We show here that the perturbation due to a plastic event occurring 
in the vicinity of a wall is more rapidly damped 
than the perturbation due to an elastic event occurring in the bulk.

\begin{figure}[t]
\includegraphics[width=7cm]{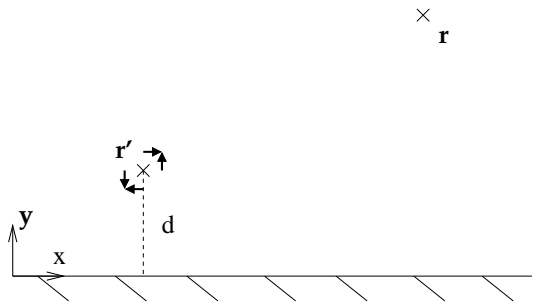}
\caption{The plastic event occurs at the vicinity of the wall : $\vert{\bf r - r'} \vert \gg d$. The quadrupolar symmetry of the plastic event is assumed not to be affected by the presence of the wall. }
\label{fig : wallevent}
\end{figure}

To be precise, a plastic event is considered to occur in the vicinity of the
bottom wall (located at $y_w = -H/2$) if it occurs at a position ${\bf r'}$ much closer to the wall than the point ${\bf r} $ where the perturbation is calculated. This condition requires $ \vert H/2+ y' \vert \ll \vert {\bf r -r'} \vert$, with the notations defined on Figure \ref{fig : wallevent}. 
We also focus here on situations where the second (top) wall is too far to play a role, that is $ \vert {\bf r -r'} \vert \ll \vert  {y' -H/2} \vert $, so that our approach in this subsection practically deals with a semi-infinite geometry.

Again, the effect of a plastic event is equivalent to the sum of that
of the four forces represented on Fig. \ref{fig : wallevent}.
Hence, we first study the displacement field induced by 
a single force ${\bf F}$ at position ${\bf r'}$ in the vicinity of the 
wall.

\begin{figure}[h]
\includegraphics[width=8cm]{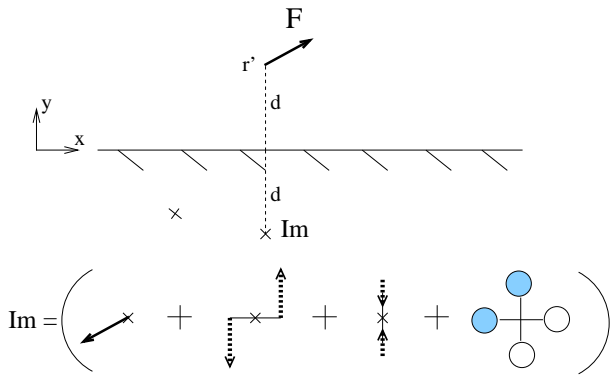}
\caption{The displacement field due to a punctual force ${\bf F}$
in a semi infinite plane at a position ${\bf r'}$ is 
 the sum of the displacement field the same force would induce at ${\bf r'}$
in an infinite medium, and that induced by its {\it image } with respect to the wall also in an infinite medium. The image consists of a punctual force ${\bf -F}$, 
two dipoles of forces of strength $2dF_x$ and $-2dF_y$ respectively, and two dipoles of potential. The resulting asymptotic long range behaviour 
is that of a quadrupole of strength $2dF_x$.} 
\label{fig : wallforce}
\end{figure}

The displacement field due to a punctual force ${\bf F}$ at ${\bf r'}$ in a semi infinite medium is the sum of the displacement field due to the punctual force and an {\it image} with respect to the wall in an infinite medium. This complex image is 
such that the displacement field it induces exactly cancel out on the wall the
displacement generated by the punctual force. Its structure is depicted 
in Fig. \ref{fig : wallforce} and recalled below \cite{pozri}. 
The location of the image is the symmetric of the pole ${\bf r'}$ with respect to the wall.  This image is the sum of a punctual force $ -{\bf F}$, two dipoles of forces $ (2d {\bf e_x},F_x {\bf e_y})$  and ($2d {\bf e_y},F_y {\bf e_y}$), and two dipoles of potential ($2d {\bf e_x},F_x d / \mu $) and ($2d {\bf e_y},-F_y d / \mu$)  where $ d =\vert y' +H/2 \vert$ is the distance between the event and the wall. 
The analytical expression for the corresponding Green function is given in \cite{pozri}.  
Its long-range behaviour, $\vert {\bf r -r'} \vert \gg d $, 
is dictated by the dipoles of forces in Figure 6, yielding
a displacement field
that scales a $\frac{1}{\vert {\bf r -r'} \vert}$.

\begin{figure}[h!]
\includegraphics[width=8cm]{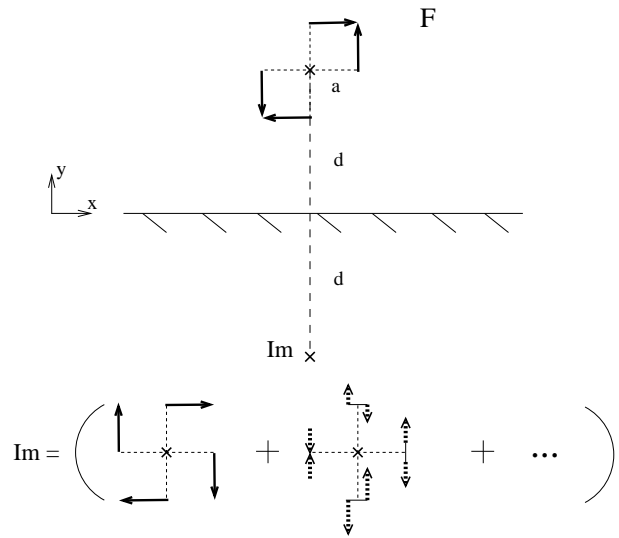}
\caption{The displacement field due to two force dipoles in a semi infinite plane
 is the sum of (i) the displacement field it would induce in an infinite medium 
and (ii) the displacement field due to its {\it image } with respect to the wall.
This image can be constructed by summing those of the four forces.
The contributions represented here (not at scale) 
are the direct images (solid arrows,
all of amplitude $F$) and
the induced force dipoles (dashed arrows, of amplitude $2(d-a)F$ and $2(d+a)F$
for the top and bottom one, of amplitude $2dF$ for the two others).
Inspection shows 
that the net dominant long-range behaviour is that of a quadrupole
twice as strong as the original one}
\label{fig : wallquadrup}
\end{figure}

\begin{figure}[h!]
\includegraphics[width=7cm]{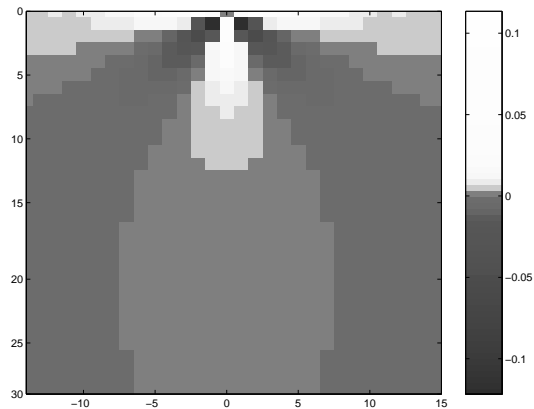}
\caption{Perturbation of the shear stress field for a plastic event (plastic deformation of amplitude $\epsilon_0=1$) occurring next to the top wall.
}
\label{prop_wall}
\end{figure}

Returning to the effect of a plastic event, 
where the source is now a set of dipoles as in Figure 2,
one could expect a similar cancellation of the first order terms, 
and thus a far-field displacement
scaling as $\frac{1}{\vert {\bf r -r'} \vert ^2}$. 
However inspection of the structure of the image show that such is not the case
(see Figure \ref{fig : wallquadrup}): the dipole of horizontal forces
is duplicated in its direct image, and the set of dipoles of vertical 
forces it generates yield a dipole of strength $4aF$. The contributions
of the (original) dipole of vertical forces are weaker due to cancelations.
Altogether one is left with a dominant term that has the
same geometry than in an infinite medium (and a 
similar $1/r$ decay for the displacement) with an amplitude twice as strong.
We have analytically checked that the stress decay
is consistent with the above picture, as for a localized event 
close to the wall we obtain a propagator that is directly related
to its equivalent in the absence of the wall Eq.  (\ref{g4p2d}) :
\begin{equation}
G_{\rm wall}(r,\theta) =  2 G^{\infty}(r,\theta) (1+ O(d/r))
\label{eq_gwall}
\end{equation}
where clearly $x-x' =r \cos(\theta)$ , $y-y'=r \sin(\theta)$.
This picture is also consistent with numerical calculations
of the propagator for an event next to the wall in a finite 
geometry, which yields the picture in 
 Fig. \ref{prop_wall} (calculation to be described in subsection IVB below).

\subsection{Propagator in a finite thickness medium}
\label{subsec_finite}

In this subsection, we turn to a medium of finite thickness $H$.
We focus again on an imposed strain situation, and therefore
on the problem corresponding to the system of equation (\ref{source_pl})
together with the no displacement boundary conditions (\ref{bc_finite}).
We indicate ways of calculating the propagator in this geometry
but mostly emphasize consequences of a single event on integral quantities.

\subsubsection{Finite $H$, Infinite $L$}

A first method to treat the case of a medium of finite thickness
and infinite length, consists in the systematic construction of
a series of images so as to cancel the displacements on both walls.
  
In the previous subsection we reported the 
explicit structure of the image of a punctual force 
needed to cancel the displacement due to this force on one of the walls.  
The image with respect to the wall at $H/2$, unfortunately 
creates a displacement at the wall $-H/2$, so that its own image with respect to the wall $-H/2$  has to be considered, and this process has to be repeated for every image. Thus formally two infinite sums of images are required to express the displacement for a punctual force  within two walls.
Following this strategy, Pozrikidis \cite{pozri} performed a full analytical 
calculation of the deformation field induced by a point force. 

Formally, the perturbation due to a plastic event can then be deduced by adding up the consequences of each of the four forces
it consists of. This leads to expressions that although exact are 
heavy to deal with and somewhat opaque. 
We therefore turn to other methods in the following, focusing on a geometry
periodic in the $x$ direction.

\subsubsection{Finite $H$, $L$ Periodic : first method}

\begin{figure}[t]
\includegraphics[width=8cm]{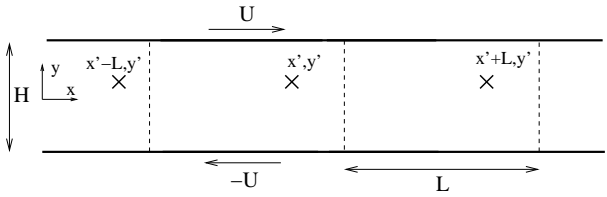}
\caption{A periodic array of plastic events at position $x' + kL, y'$.}
\label{fig:cisperiodic}
\end{figure}

We now focus on a system of thickness $H$
and of finite extent $L$ in the $x$ direction,
and consider periodic boundary conditions in that direction.
This is equivalent to analyzing in an unbounded geometry 
the effect of a periodic array of 
plastic events of a given amplitude at positions $(x' + k L,y')$ with $k \in \mathbb{Z}$ 
(Figure \ref{fig:cisperiodic}),
or formally $\hat{\epsilon}_{xy}^{pl}({\bm r}) = a^2 \epsilon_0 
\sum_{k \in \mathbb{Z}}\delta({\bf r}-({\bm r'}+kL{\bm x}))$. 
The resulting displacement field  ${\bf u}^1$ can then be viewed
as the sum of the displacement field induced by a similar
periodic array of plastic events {\em in an infinitely thick medium} ${\bf u}^{\infty }$
and 
of a correction term ${\bf v} ^ {H}$ due to the finite size $H$. 
${\bf u}^1={\bf u}^{\infty }+{\bf v} ^ {H}$ 
is a function of $x-x',y \mbox{\, and \,} y'$. 

The displacement field ${\bf u}^{\infty }$ induced by 
the periodic array of plastic events in an infinite medium can be 
expressed using the propagator for a single event (\ref{propag_inf}) :
\begin{equation}
{\bf u}^{\infty }(x,y) =a^2 \epsilon_0 \sum_{k \in \mathbb{Z}} {\bf P}^{\infty}(x-x'-k L,y') 
\end{equation}
Given its periodicity in the $x$ direction, this
expression can be formally decomposed in Fourier series:
\begin{eqnarray}
{\bf u}^{\infty }(x,y)={\bf U}_0^{\infty} (y)  + \sum_{n=1}^{\infty}& 
(\, {\bf U}_{cn}^{\infty}(y) \cos(2 n \pi x/L) \nonumber \\ 
+  &{\bf U}_{sn}^{\infty}(y) \sin(2 n \pi x/L) \,)\nonumber \\
\end{eqnarray}
From equation (\ref{eq_sum}), the components of the zeroth-mode
vector function are:
\begin{eqnarray}
\label{1stterm_inf}
U^{\infty}_{0x}(y) &=&  \mbox{\,Sign\,}(y-y') a^2 \epsilon_0 /L  \nonumber \\
U^{\infty}_{0y}(y) &=& 0
\end{eqnarray} 

The correction displacement field ${\bf v} ^ {H}$ is the solution 
of incompressible linear elasticity with no source
(i.e. the set of equations (\ref{elasticity},\ref{mech_eq}) with no plasticity) 
but with the boundary conditions required to grant that ${\bf u}^1$ is zero on the walls:
\begin{equation}
{\bf v} ^ {H}(x, \pm H/2) + {\bf u}^{\infty }(x,\pm H/2) =0
\label{eqbccorr}
\end{equation}
${\bf v}^H$ is thus also periodic and can be written : 
\begin{eqnarray}
{\bf v}^{H}(x,y) ={\bf V}_0^{H} (y) 
+  \sum_{n= 1}^{\infty}  &(\,{\bf V}_{cn}^{H}(y)& \cos(2 n \pi x/L)\nonumber \\ 
 + & {\bf V}_{sn}^{H}(y)& \sin(2 n \pi x/L) \,)\nonumber
\end{eqnarray}
where the functions ${\bf V}_{cn}^{H}$, ${\bf V}_{sn}^{H}$ can be 
calculated independently (i.e mode by mode)
with the boundary conditions: ${\bf V}_{cn} ^ {H}(\pm H/2) = - {\bf U}_{cn}^{\infty }(\pm H/2)$, ${\bf V}_{sn} ^ {H}(\pm H/2) = - {\bf U}_{sn}^{\infty }(\pm H/2)$.
We skip the full calculation in this subsection
(which can be performed e.g. as in the low Reynolds number hydrodynamic
study in \cite{Ajd}), as we display in the next subsection
an exact expression of the displacement calculated 
in a framework that is more amenable to numerical simulation. 

Instead we focus here on the zeroth mode ${\bf V}_{0}^{H}(y)$. 
From equations (\ref{1stterm_inf}) and (\ref{eqbccorr}),
its components are simply:
\begin{eqnarray}
V^{H}_{0x}(y) &=& - 2 a^2 \epsilon_0 \frac{y}{HL}  \nonumber \\
 {V}^{H}_{0y}(y)&=& 0 .
\end{eqnarray} 

This suffices to deduce consequences of the plastic event in terms
of integrals over constant $y$ lines: 
\begin{equation}
\int^{L/2}_{-L/2} {\bf u}^1 dx = L  ({\bf V}_0^{H} (y) + {\bf U}_0^{\infty} (y) )\nonumber 
\end{equation}
so that
\begin{eqnarray}
\int^{L/2}_{-L/2}  u^1_x(x,y) dx &=&- 2(a^2 \epsilon_0 \frac{y}{H}) +  \mbox{Sign}(y-y') a^2 \epsilon_0  \nonumber \\
\int^{L/2}_{-L/2}  u^1_y(x,y)  dx &=& 0 \nonumber
\end{eqnarray}

Then, linear elasticity implies that the overall variation of the force on the top
plate due to the plastic event is:
\begin{equation}
\label{sigma_sum}
\delta F = \int^{L/2}_{-L/2} {\sigma_{xy}}^1 dx =- 2  \frac{\mu \epsilon_0 a^2}{H}
\end{equation}
The corresponding drop of the average shear stress in the medium is obviously
$\delta \langle\sigma\rangle = \delta F/L$.

This exact expression shows that {\em a single plastic event 
(of a given fixed amplitude $\epsilon_0 a^2$) 
results in the release of the net force exerted by the medium on the walls by a quantity scaling as $\frac{1}{H}$. Remarkably, this quantity is independent on the position of the event in the medium.} A corollary is that a finite density
of plastic events $\phi$, will relax this total force by an amount $\sim (\phi HL) \delta F$, corresponding to a relaxation of the average stress 
$\delta \langle\sigma\rangle =\delta F/L$ independent of the size of the system.

Note that the integrals over a period $L$ calculated above 
for the consequences of a periodic array of plastic events,
are equal to the integrals over $x$ from $-\infty$ to $+\infty$
in a finite $H$ infinite $L$ geometry for the consequences of a single 
plastic event.  
For example, in the latter geometry 
$\delta F = \int^{\infty}_{-\infty} {\sigma_{xy}}^1 dx =- 2  \frac{\mu \epsilon_0 a^2}{H}$, 
which clarifies in what sense
equation (\ref{eq_sum}) obtained in the previous section for an infinite
medium corresponds to the limit $H \rightarrow \infty$.
The global relaxation due to a single plastic event is consequently
directly related to the finite size of the system.  

An alternative derivation of this finite-size dependence
of the force (or average stress) relaxation
is proposed in appendix \ref{sec_schemaphys}, 
which yields a somewhat complementary physical insight.

\subsubsection{Finite $H$, $L$ Periodic : second method}

Although we aim to focus in this paper on the qualitative aspects
presented previously, we propose here an explicit formula 
for the stress perturbation induced by a localized plastic event in a finite 
size geometry.
The actual expression is presented in reciprocal space, which may 
appear at first somewhat cumbersome, but turns out to be convenient 
in numerical calculations. For sake of readability, we provide here only
the principles of this two step derivation and the 
resulting formulae, the details of the calculations being
presented in appendix \ref{sec_cal_wall}.\\

First step: we formally extend the actual system 
($0<y<H$, and $L$ periodic along $x$)
by the following two operations: first an antisymmetric image system
is constructed that extends to $-H<y<0$, then the $2H$ thick
resulting system is repeated periodically in the $y$ direction. 
If the plastic strain in the original system is
$\epsilon^{pl}(x,y)$, our construction 
yields a system without walls that is $2H$ periodic in the $y$ direction,
with in the upper half period $H>y>0$
a plastic strain $\epsilon^{pl *} (x,y) = \epsilon^{pl}(x,y)$, 
and in the lower half 
$-H<y<0$ an {\it antisymmetric image} plastic strain $ \epsilon^{pl *} (x,y) = - \epsilon ^{pl}(x,-y)$. 
This construction ensures that the overall displacement 
generated by these strains
has an $x$ component that is symmetric by reflection by the planes 
$y=0$ and $y=H$, and an $y$ component that is antisymmetric in the same operations. 
We have therefore generated a solution that satisfies 
the condition of a zero 
$y$ component of the displacement on the $y=0$ and $y=H$ planes 
(loci of the walls in the original system).

Second step: We now want to cancel the remaining displacements along $x$ without 
modifying the above result, and without adding sources in the system $0<y<H$.
This can be achieved by adding on the planes $y=0$ and $y=H$ appropriate force
fields $f_x$ directed along $x$ (again asymmetric and $2H$ periodic along $y$).
Given the symmetry and periodicity of the system it is clear that the $y$ displacement on the walls is not modified by this addition.

\begin{figure}[t]
\includegraphics[width=7cm]{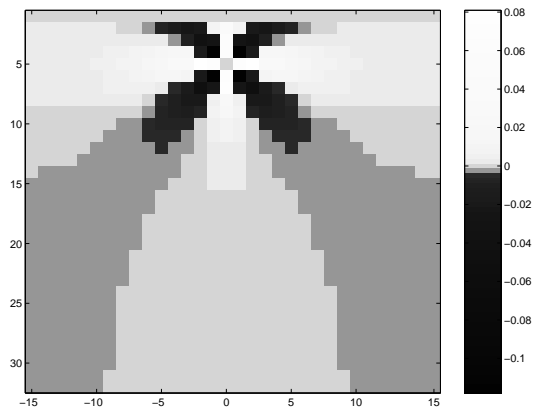}
\caption{Perturbation of the shear stress field for a plastic event (plastic deformation of amplitude $\epsilon_0 a^2=1$) occurring in the vicinity the wall.
The discretization corresponds to $H/a=L/a=32$ 
}
\label{prop_wallint}
\end{figure}

When this is achieved, we have in the upper half $0<y<H$,
a solution to (\ref{source_pl}) that satisfies the
no displacement boundary condition on the walls (\ref{bc_finite}). 
The corresponding stress field can be expressed in Fourier series :
\begin{equation} 
\sigma^1(x,y) =\sum_{m,n\in\cal{Z}}  e^{i p_m x}   e^{i q_n y} \hat{\sigma}^*(m,n) \nonumber \\
\end{equation}
with  $p_m =\frac {2 \pi m }{L}$ and $q_n = \frac{2 \pi n} {2H}$.
It is the sum of the term directly generated by the plastic strains
${\hat \epsilon}^*$ and that due to the added force fields:
\begin{eqnarray} 
\hat{\sigma}^*(m,n) &=& 2 \mu \lbrace \hat{G}^{\infty}(m,n)  \hat{\epsilon}^{pl*} (m,n) \nonumber \\ &+& \frac{1}{2}({\it i}p_m \hat{O}_{xy}(m,n)+ {\it i}q_n \hat{O}_{xx}(m,n)) \hat{f}_x(m,n)
\rbrace
\nonumber \\
\hat{\sigma}^*(0,0)&=& 0
\label{eq_sig_four} 
\end{eqnarray}
The propagator $G^{\infty}$ and the Oseen tensor $\bf{O}$ are the 
direct counterparts for periodic systems of those 
defined in section \ref{sec_infinite}. Of course the force field $f_x$ 
in the above expression is itself proportional to the plastic strain;
the corresponding formulae are given in appendix \ref{sec_cal_wall}.

Inverting back to real space allows to compute numerically
the response to a localized event. We have checked that
for a large system the stress created by a plastic event far from the walls
according to this formula is close to that 
obtained analytically in section III for an infinite medium.
Similarly, for an event directly neighbour to the wall, we recover the results of
subsection IVA for a semi-infinite medium. 
Fig. \ref{prop_wallint} represents an intermediate situation : the event occurs in the vicinity of the wall.

\section{Conclusions and perspectives}
\label{sec_conseq}

Starting from a general elasto-plastic model, 
we have computed in different 2D geometries
the modification of the shear stress resulting from a localized 
plastic event with a symmetry of simple shear. 
We have first calculated the corresponding perturbation 
in an infinite system forced with a symmetry of shear. The stress field is of quadrupolar symmetry and decreases with a power law $\frac{1}{r^2}$ in two dimensions. Then, we showed that the stress field perturbation due to a plastic event occurring close to a wall has a modified near field 
structure but decays far away with the same law and pattern, 
although with an amplitude twice as large.
Eventually, we have proposed two ways 
of calculating the perturbation of the stress field 
due to a plastic event occurring in a finite medium. The first one allowed us to demonstrate 
in a simple way that a plastic event of a given strain amplitude relaxes 
the average stress by an amount which is independent of its position (i.e. distance to the walls) and inversely proportional to the size of the system. The second one 
allowed us to derive explicit expressions that permits calculation of the whole
stress field in a finite-size geometry. 

The extension of our results to a three dimensional situation
is rather straightforward, and the qualitative statements are obviously similar.
Also we have focused on a scalar description of plastic strain and
induced stress, but the same steps can be taken if one seeks
to describe the whole tensorial stress field generated
by  plastic events of arbitrary symmetry. 
Eventually, we have focused in section IV on a situation where the walls 
were kept fixed during the plastic event. The extension of our results to 
situations where the force on the plates is kept fixed is immediate:
one simply needs to add to our solution 
a simple shear displacement (corresponding to a stress
$-\delta F/L$ with the notations of section IV). 

This study is meant to be a first step towards the modeling of the flow
of an elasto-plastic material. The next one consists
in plugging in a plastic law that describes the onset and evolution 
of the localized plastic events.  Coupling such a local plastic behavior to the long-range elasticity described here should yield 
interesting collective behaviours and hopefully insights in the flow mechanisms.
For such an endeavour our quantification of finite-size effects is important
for the steady-state average balance between the stress released by the plastic events and that imposed by the elastic loading.
Also the geometry and decay law of the elastic perturbation is to be 
considered when addressing the emergence of a collective/cooperative organization
at low shear rates.  Important questions regarding possible spatial and temporal
heterogeneities in such flows (\cite{pignon,varnik,cous,deb,kabla} 
could be addressed at the light of the present results:\\
- We have shown that a localized plastic event relaxes the average stress but
also modifies the stress pattern on {\em all} the lines of the system
decreasing the stress of some elements but increasing that of others. 
Does this allow shear banding with a limited zone flowing 
in coexistence with a non flowing region ?\\  
- The average stress released during an event 
of given strain amplitude is independent of the position of the event 
in the medium, and is the same on all lines. 
Yet the geometry of the propagator 
is clearly modified by the proximity of a wall.
Do these elements favor a localization of the flow at the wall ?\\
These questions will be addressed in a forthcoming publication.



\appendix
\section{Finite size effects.}
\label{sec_schemaphys}

We propose an alternative simple approach which gives a different
physical insight into 
the effects of finite size evidenced subsection \ref{subsec_finite}. 
The following 
argument holds equally for an infinite $L\rightarrow \infty$ geometry or 
for an $L$ periodic medium, and is here presented for the latter.
We consider a plastic event 
occurring at $(x',y')$ described by 
$\epsilon^{pl}_{xy}(x,y)=\epsilon_o a^2 \delta (x-x',y-y')$ (and its repeated
images along $x$ due to the $L$ periodicity). The stress perturbation is:
\begin{equation}
\sigma_{xy}^1(x,y)=
\int^{L/2}_{-L/2}\!\!\!\!\!\!\!\!dx_1
\int^{H/2}_{-H/2}\!\!\!\!\!\!\!\!dy_1
\,G^{HL}(x-x_1,y,y_1)\epsilon_{xy}^{pl}(x_1,y_1)
\end{equation}
which actually defines the propagator $G^{HL}$ for the 
present periodic medium. The corresponding 
force release on the bottom of a layer at height $y$ is
\begin{equation}
\delta F = \int^{L/2}_{-L/2} \!\!\!\!\!\!dx\,\, 
\sigma_{xy}^1(x,y) = a^2 \epsilon_0 
\int^{L/2}_{-L/2}\!\!\!\!\!\!dx\,\,  G^{HL}(x-x',y,y')
\end{equation}
We remark that this can be rewritten
\begin{equation}
\delta F
=
\int^{L/2}_{-L/2}\!\!\!\!\!\!\!\!dx_1
\int^{H/2}_{-H/2}\!\!\!\!\!\!\!\!dy_1
\,G^{HL}(x-x_1,y,y_1)(\epsilon_0a^2 \delta(y_1-y'))
\end{equation}
which means (see (A1)) that the variation of the integrated stress over a layer 
at height $y$ ($\delta F$) induced by an event occurring at position $x',y'$, is equal to the stress variation at a site $(x',y)$ induced by a continuous 
sum of events on a layer $y'$. 

\begin{figure}[t]
\includegraphics[width=8cm]{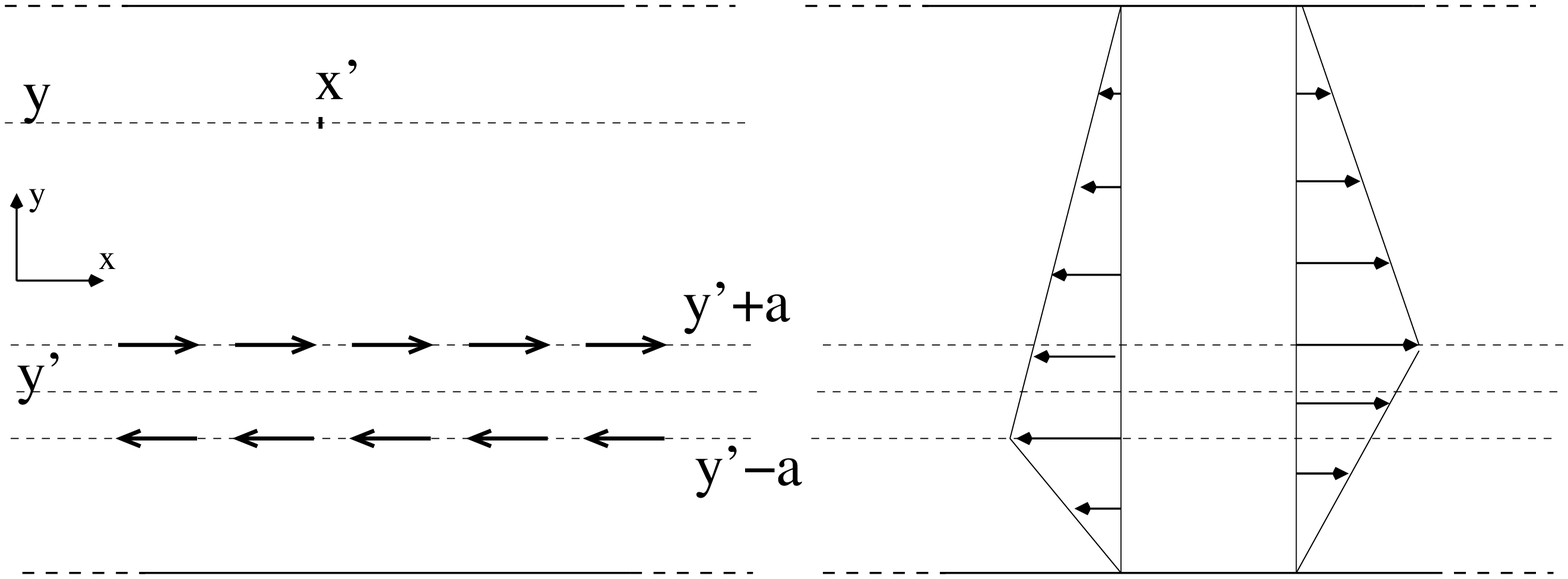}
\caption{Left: a continuous line of events at height $y'$ 
is equivalent, using the representation of Figure 2 for a single event, 
to two lines of horizontal forces at $y'+a$ and $y'-a$.
Right: displacement profiles resulting from the sum of negative forces 
at $y'-a$ (left), and from the sum of positive forces at height $y'+a$ (right).}
\label{sum2p}
\end{figure}

Representing these events by pairs of dipoles,
the sum reduces to that of the $x$ components as the $y$ components cancel out
upon summation on the line, so that we are looking for the stress generated
at $(x',y)$ by the two lines of forces at $y'-a$ and $y'+a$ depicted on 
Figure \ref{sum2p}. 
The displacement field resulting from each of these sums separately
are easily calculated and depicted on the right of Figure \ref{sum2p}. 
For example the second one is the solution of the following problem:
on each side of the plane $y=y'-a$, the medium is purely elastic, with no displacement at the walls, and at  $y=y'-a$  
a continuous displacement and a jump in the stress of amplitude $\mu \epsilon_0 a$.
  
\begin{figure}[h]
\includegraphics[width=8cm]{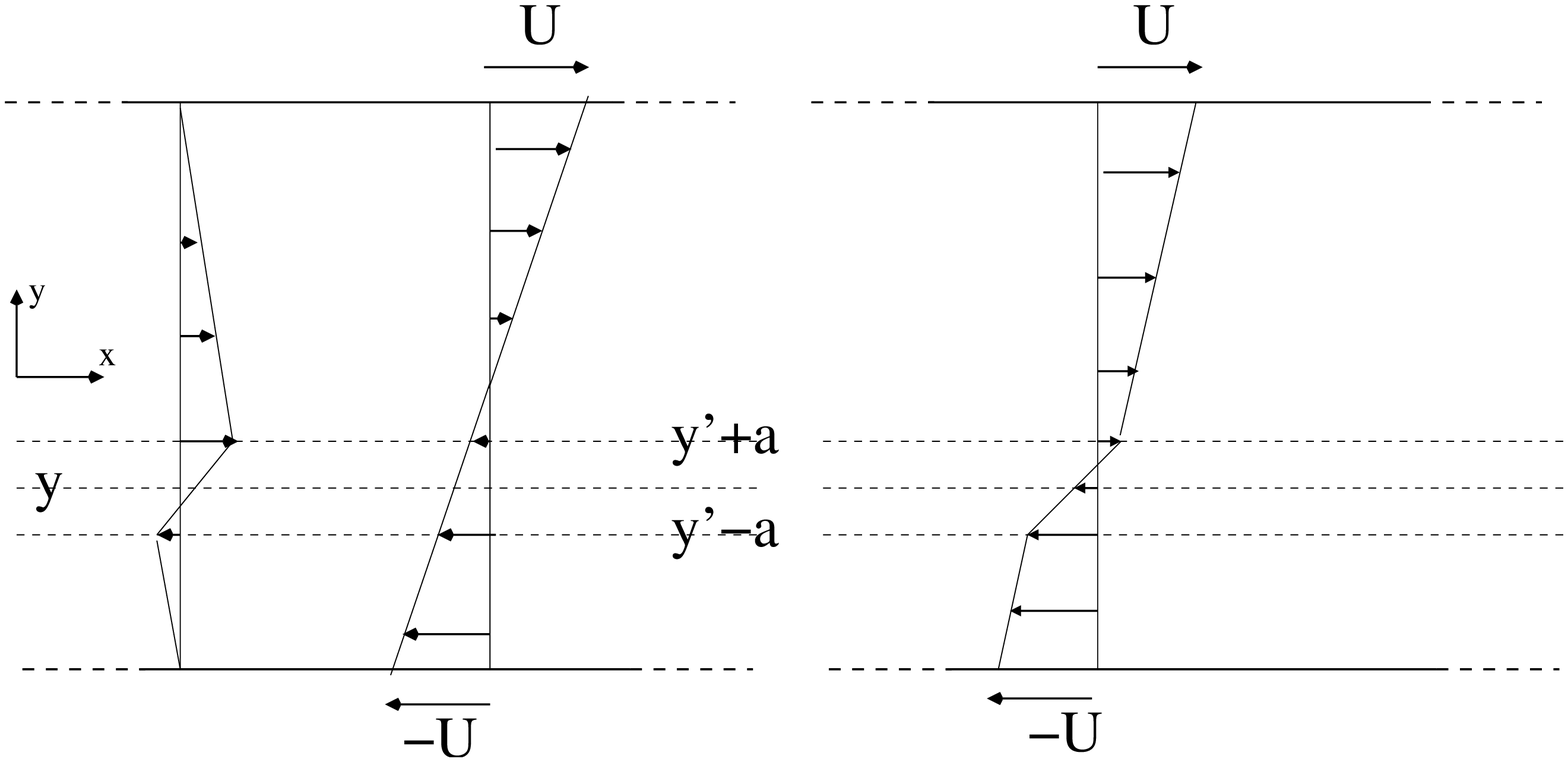}
\caption{Left : displacement profile due to the line
of events at $y'$ (obtained by summing the two profiles of Fig. \ref{sum2p}
right), and schematic description of the elastic loading.
Right:
total response to the global forcing and a line of plastic event (a homogeneous fracture), 
obtained by summing the two profiles on the left.}
\label{velpro_s1}
\end{figure}

The total effect of the continuous line of events is the displacement 
field presented on the left of Figure \ref{velpro_s1}.
The slopes of the profile for $y > y'+a$, and $y < y'-a $  are the same,
and correspond to $\delta F=-2\frac{\mu\epsilon_0a^2}{H}$,
as derived differently in the main text, 
see equation (\ref{sigma_sum}). Remarkably, this value,
which corresponds to the force release on the top wall,
is independent of the position of the plastic event.

We remark that the deformation profile obtained here is very similar
to the one obtained by Kabla and Debr\'egeas in their 
explicit simulation of a shear foam \cite{kabla} 
(which in their case was the line average 
consequences of a single local event, equivalent to the present local consequence
of a line of events according to the argument below (A3)).
 
As a follow-up along the same picture, the cumulative displacement
after loading plus a continuous line of plastic 
events is represented on the right of
Figure \ref{velpro_s1} :  the total shear in the system
is then the homogeneous shear due to the forcing
minus the shear release due to the plastic events. 


\section{Stress field induced 
by a plastic deformation in a finite medium}
\label{sec_cal_wall}

In this appendix we provide the explicit derivation
of the stress field change due to a plastic activity in 
a finite system, follow the strategy outlined in subsection IVB.3.
The aim is to obtain the response of the system depicted
on figure \ref{fig:cisperiodic} to a strain solicitation 
$\epsilon^{pl}(x,y)$ with "stick" boundary conditions on the wall
similar to \ref{bc_finite}. For this calculation
we choose to take the origin of the $y$ axis on the bottom
wall so that the walls (and thus the boundary conditions)
are at $y=0$ and $y=H$.

\begin{figure}[h]
\includegraphics[width=7cm]{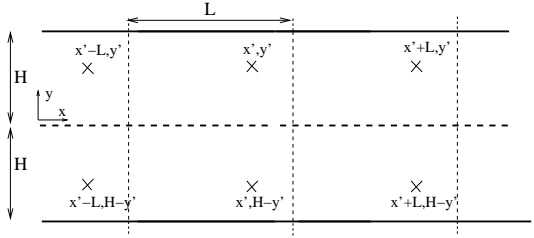}
\caption{Auxiliary system corresponding to the real system of figure
\ref{fig:cisperiodic}:
it is periodic in both direction and symmetric in the $y$ direction. The 
driving plastic strain field $\epsilon^{pl*}$ is antisymmetric.}
\label{fig_dbleperiod}
\end{figure}

\paragraph{{\it Geometry of the auxiliary system-}}
To solve this problem, an auxiliary system with no walls
is constructed. Its geometry is that of the 
original system, symmetrized with respect to $y=0$, 
and periodized along $y$. The new system 
has thus a period $L$ in the $x$ direction, 
and a period $2H$ in the $y$ direction, see Fig. \ref{fig_dbleperiod}.
Fields in this new system are described by stars.
The symmetry with respect to the plane $y=0$ is made in such a way
that the plastic strain is antisymmetric 
$\epsilon^{pl *}(x,-y)=-\epsilon^{pl *}(x,y)$,
which is equivalent in the force dipole representations to having 
the force fields in the lower half to be the symmetric of that in the upper
part $\bm{F}^*(x,y)= \bm{F}^*(x,-y)$. 
The symmetry and periodicity of the
problem ensures that the displacement field $\bf{u}^{*}$ 
generated by this plastic strain satisfies 
verify by symmetry and periodicity :
\begin{equation}
u_y ^* (x,0) =0 ; u_y ^* (x,H) =0 
\end{equation}

\paragraph{{\it Response to a plastic event-}}
This new elastic system is submitted to a 
plastic strain $ \epsilon^{pl *}(x,y) = \epsilon^{pl}(x,y) \mbox{\, if \,} y > 0$, and $ \epsilon^{pl *}(x,y) = - \epsilon^{pl}(x,-y) \mbox{ \, if \,} y < 0$. 
For reasons explained in the text (section IVB.3), a force field $f_x^*$ 
is added at the location of the walls in the real system,
to cancel the displacements $u_x(x,0)$ and $u_x(x,H)$ created by the plastic strains.
\begin{equation}
f_x^*(x,y)=f_0^*(x)\delta(y) + f_H^*(x) \delta(y-H)
\end{equation} 
This "no displacement on planes $y=0$ and $y=H$" condition
is implemented in reciprocal space. 
We introduce notations for the Fourier series:
\begin{eqnarray}
\bm{u}^*(x,y) &=&\sum_{m,n\in\cal{Z}}  e^{i\frac{2 \pi m x}{L}}   e^{i\frac{2 \pi n y}{2H}} \hat{\bm{u}}^*(m,n) \nonumber \\
\hat{\bm{u}}^*(m,n) &=& \frac{1}{2HL}  \int \! dx dy\, \bm{u}^*(x,y)\, 
e^{-i\frac{2 \pi m x}{L}} e^{-i\frac{2 \pi n y}{2H}} \nonumber \\
p_m =\frac {2 \pi m }{L} &;& q_n = \frac{2 \pi n} {2H} ;\, q^2= p_m^2+ q_n^2 \nonumber \\
\end{eqnarray}
From our calculations for an infinite medium in section \ref{sec_infinite}, it is 
logical to write the total displacement using propagators for plastic events and 
the Oseen propagator for simple forces:
\begin{eqnarray}
\hat{\bm{u}}^*(m,n) &=& \hat{\bm{P}}^{\infty}(m,n) \hat{\epsilon}^{pl*} (m,n) \nonumber \\ &+& \hat{\bm{O}}(m,n) \hat{f}_x(m,n)
\end{eqnarray}
The equations for $\bm{P}^{\infty}$ and the Oseen tensor $\bm{O}$
in the present representation (Fourier series) are formally exactly
similar to those obtained
 in (\ref{eq_oseen}) and (\ref{propag_inf}) in terms of Fourier transforms
(with $q_x \rightarrow p_m$ and $q_y \rightarrow q_n$).

The condition $u^*(x,0)=0; u^*(x,H)=0$ allows to compute
$\hat{f}^*_x(m,n)$.
\begin{eqnarray} 
\hat{f}^*_x(m,n) &=& \frac{\sum_{1} \frac{2i}{q^4} p_n'(q_m^2-p_n'^2) \hat{\epsilon}^{pl*}(m,n')}{\sum_{1} \frac{p_n'^2}{\mu q^4}}, 
\nonumber \\
  &&\mbox{for odd n}\nonumber \\
\hat{f}^*_x(m,n) &=& \frac{\sum_{2} \frac{2i}{q^4} p_n'(q_m^2-p_n'^2) \hat{\epsilon}^{pl*}(m,n')}{\sum_{2} \frac{p_n'^2}{\mu q^4}},
\nonumber \\
 && \mbox{for even n and \,} m \neq 0 \nonumber \\
\hat{f}^*_x(0,n) &=& 0 \mbox{ \, for even n} \nonumber
\end{eqnarray}
$\sum_1$  describes the sum over odd values of $n'$, 
and $\sum_2$ the sum over even values of $n'$. \\
The shear stress in this geometry can then be derived from the displacement field:
\begin{equation}
\hat{\sigma}^*(m,n) = 2 \mu (\frac{1}{2}({\it i}p_m \hat{u}^*_y(m,n) + {\it i}q_n \hat{u}^*_x(m,n) - 
\hat{\epsilon}^{pl*}(m,n))
\end{equation}
For symmetry reasons, $\hat{\sigma*}(0,0)=0$.

\paragraph{Back to the real system -} The displacement field $\bf{u}^*(x,y)$
that we have constructed and computed 
satisfies the elasticity equations for $H>y>0$
with the plastic strain $\epsilon^{pl}(x,y)$ as only source in that area,
and verifies the boundary conditions $u^*(x,0)=0$ and $u^*(x,H)=0$.
It is therefore the solution to our initial problem for the real system:
for $H>y >0$, $\bf{u}^1(x,y) = \bf{u}^*(x,y)$,
and $\sigma^1(x,y) = \sigma^*(x,y)$. 
Hence, the expression of the stress as given in Equation (\ref{eq_sig_four}).

\end{document}